\documentclass[aps, prb,amsmath,amssymb,reprint,floatfix,superscriptaddress,]{revtex4-1}

\pdfoutput=1 

\usepackage{graphicx}
\usepackage{dcolumn}
\usepackage{bm}

\begin{document}

\title{Extremely high negative electron affinity of diamond via magnesium adsorption}

\author{K. M. O'Donnell}
\email{kane.odonnell@curtin.edu.au}
\affiliation{Department of Physics, Astronomy and Medical Radiation Science, Curtin University, Bentley, Western Australia 6102, Australia}
\author{M. T. Edmonds}%
\affiliation{School of Physics, Monash University, Clayton, Victoria 3800, Australia}
\author{A. Tadich}
\affiliation{Australian Synchrotron, 800 Blackburn Road, Clayton, Victoria 3168, Australia}
\author{L. Thomsen}
\affiliation{Australian Synchrotron, 800 Blackburn Road, Clayton, Victoria 3168, Australia}
\author{A. Stacey}
\affiliation{School of Physics, University of Melbourne, Parkville, Melbourne Victoria 3010, Australia}
\author{A. Schenk}
\affiliation{Department of Physics, La Trobe University, Bundoora, Victoria 3086, Australia}
\author{C. I. Pakes}
\affiliation{Department of Physics, La Trobe University, Bundoora, Victoria 3086, Australia}
\author{L. Ley}
\affiliation{Lehrstuhl f\"{u}r Laserphysik, Friedrich-Alexander-Universit\"{a}t Erlangen-N\"{u}rnberg, Staudtstr. 1, 91058 Erlangen, Germany}
\affiliation{Department of Physics, La Trobe University, Bundoora, Victoria 3086, Australia}

\date{\today}

\begin{abstract}
We report large negative electron affinity (NEA) on diamond (100) using magnesium adsorption on a previously oxygen-terminated surface. The measured NEA is up to $(-2.01\pm0.05)$ eV, the largest reported negative electron affinity to date. Despite the expected close relationship between the surface chemistry of Mg and Li species on oxygen-terminated diamond, we observe differences in the adsorption properties between the two. Most importantly, a high-temperature annealing step is not required to activate the Mg-adsorbed surface to a state of negative electron affinity. Diamond surfaces prepared by this procedure continue to possess negative electron affinity after exposure to high temperatures, air and even immersion in water.
\end{abstract}

\maketitle

\section{Introduction}

The negative electron affinity of diamond allows for barrier-free electron emission.\cite{Himpsel:1979vd} The resulting electron yield is orders of magnitude higher than materials with a positive electron affinity,\cite{ODonnell:2013el, Shih:1997bt} and hence diamond is the material of choice for many emerging electron emission applications in both vacuum\cite{Sun:2014ih, Chang:2010fa} and solution.\cite{DiZhu:2013bb} The electron affinity of diamond is controlled by the surface termination. With a suitable modification to the surface dipole, the conduction band minimum will be driven above the vacuum level at the surface, a condition of true negative electron affinity (NEA). Hydrogen termination is the standard method for achieving NEA on diamond and has been studied extensively.\cite{Cui:1998br, Graupner:1998ub, Graupner:1999up, VanderWeide:1994wz, kanetemp:2014} The downside of using hydrogen termination is that it is susceptible to both electronic and chemical degradation. Electronic degradation results from p-type surface transfer doping that takes place spontaneously in air for the NEA surface; charge transfer results in a large internal electric field and upward band-bending that imposes a barrier to electron emission notwithstanding that the condition of NEA remains.\cite{Maier:2000uq, Riedel:2004gl, Takeuchi:2003jr} A 200-fold reduction in electron yield has been observed owing to this effect.\cite{Takeuchi:2003jr} Chemical degradation results from slow oxidation of the surface in poor vacuum in the presence of an excitation source (e.g. X-rays, UV light or electrons).\cite{Foord:2001ux,Anonymous:A0ci7tur} Since oxidized diamond has a positive electron affinity, gradual oxidation results in a shift from negative to true positive electron affinity. Both electronic and chemical degradation restrict stable diamond electron emission to ultra high vacuum environments. 

To alleviate these effects, we have studied NEA diamond surfaces that are based on incorporating light metals into the oxygen-terminated diamond surface.\cite{ODonnell:2010fw, ODonnell:2013el, ODonnell:2014cp} The strategy takes advantage of the fact that diamond has no native oxide and instead can be prepared to a state of a true ``atomic'' oxygen termination. Consequently, one has very precise control over the surface dipole. Theoretical and experimental efforts initially focused on lithium as the diamond lattice constant makes heavy alkali metals like caesium unsuitable for ordered, high-coverage surface layers. We have shown that a robust negative electron affinity is induced on oxygen-terminated diamond following lithium adsorption and thermal activation (`lithiation').\cite{ODonnell:2013el} Owing to the presence of oxygen lone pair states within the bandgap, these diamond surfaces typically have a Fermi level position approximately 0.9 eV above the valence band maximum. For boron-doped diamond this leads to large downwards band-bending that prevents p-type surface transfer doping from affecting electron emission. Indeed, we observe no air-induced surface transfer doping on lithiated diamond. Consequently, lithiated diamond is a substantially more robust emitter with respect to poor vacuum environments. 

On the other hand, the activation procedure for lithiated diamond requires temperatures in excess of 600$^{\circ}$C. Although this is certainly not as aggressive as hydrogen termination (usually via hydrogen plasma at temperatures around 800$^{\circ}$C), it is inconvenient for in-situ regeneration. Hence, we have sought a system with similar surface chemistry to lithiated diamond without the need for activation. Recently a computational study was carried out comparing different alkali metals and magnesium adsorbed on oxygen-terminated diamond.\cite{ODonnell:2015iu} In that study it was found that the adsorption energies, bond angles and electron affinities predicted for half a monolayer (ML) of Mg on oxygen-terminated diamond are very similar to that for a full monolayer of Li on the same substrate. This result is unsurprising given the so-called diagonal relationship for much of the solid-state chemistry of these two elements.

Here we present the results of the first experimental study of Mg adsorption on the oxygen-terminated diamond (100) surface. Kelvin probe and photoemission measurements are used to fully determine the band alignment at the surface. Adsorption of 0.5ML Mg leads to an electron affinity of -2.0 eV, the largest ever reported NEA. We show that NEA remains after annealing, exposure to air and even after water immersion.

\section{Experimental}

Experiments were carried out on single-crystal (100) natural diamond substrates (Delaware Diamond Knives). Conductive, boron-doped overlayers were grown on the substrates via chemical vapour deposition at the Melbourne Centre for Nanofabrication. Samples were acid washed in sulphuric and nitric acid then oxygen-terminated using a 50 W oxygen plasma for five minutes at room temperature prior to introduction into the ultra-high vacuum endstation of the Soft X-ray beamline at the Australian Synchrotron.\cite{Cowie:2010cm} Typical sample surface roughness was low (Ra $<$ 1 nm, Rq $<$ 1 nm, Rmax $<$ 10 nm) as measured via atomic force microscopy (Bruker FastScan). Prior to experiments, samples were annealed overnight between 300-400$^{\circ}$C to remove all atmospheric contaminants. Magnesium was deposited using a crucible source (MBE Komponenten NTEZ) with the deposition rate monitored by a quartz crystal microbalance (QCM). The base pressure of the end station remained at ultra-high vacuum, $p < 5\times 10^{-10}$ mBar, at all times. The evaporation of Mg was not accompanied by co-deposition of contaminants as verified via core level photoemission. The contact potential difference (CPD) between the stainless steel tip of a Kelvin probe (KP Technologies) and the sample was measured immediately after deposition followed by photoemission. The CPD and secondary electron cutoff for a gold reference sample were measured after each Mg deposition to maintain accurate Kelvin probe calibration for workfunction measurements. 

\section{Results/Discussion}

\begin{figure}[htb]
	\includegraphics[width=0.75\columnwidth]{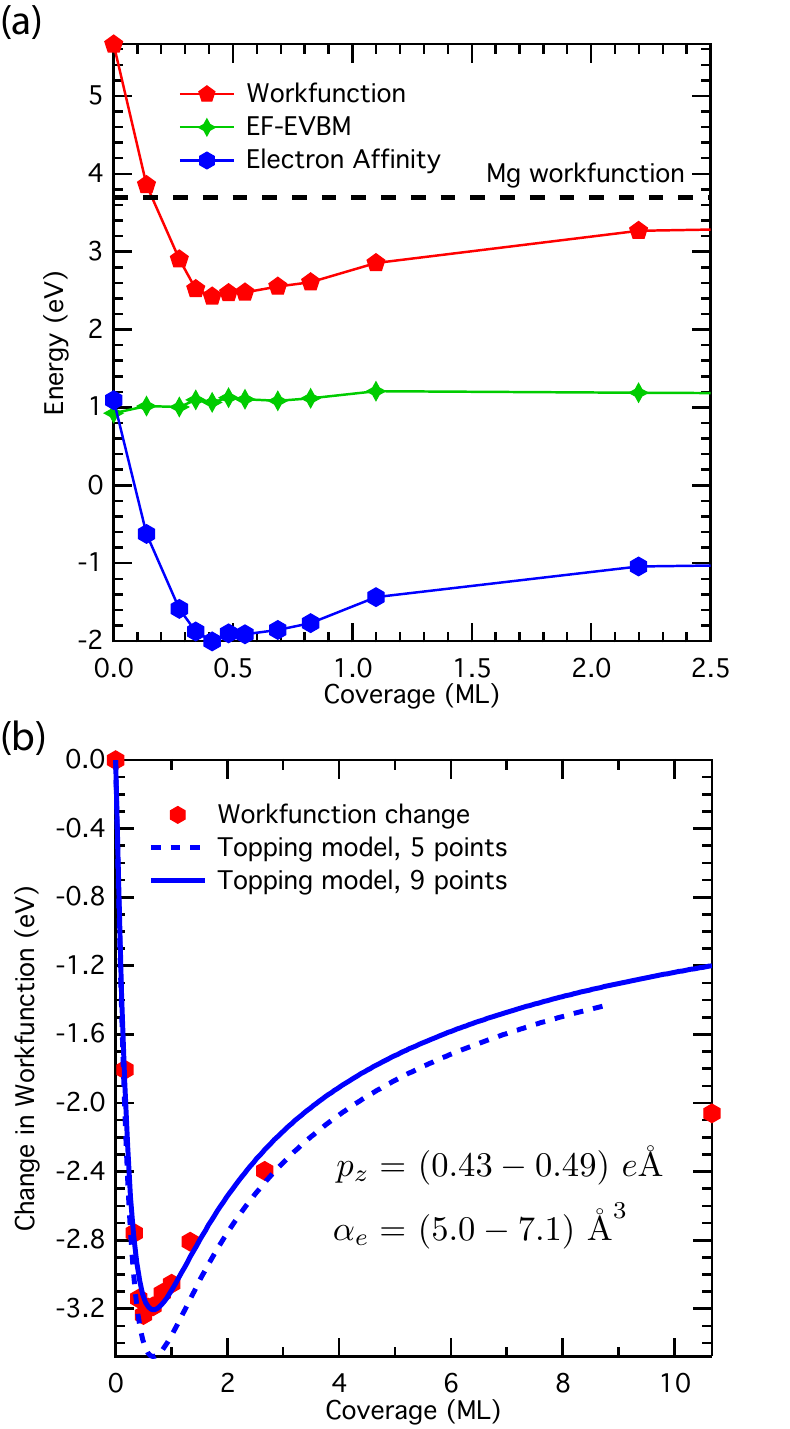}
	\caption{(a) Workfunction, electron affinity and $E_{F}-E_{VBM}$ for Mg deposition on oxygen-terminated diamond as a function of coverage. The bulk Mg workfunction is indicated. (b) Change in workfunction with respect to the clean surface fit to the Topping model from Ref \onlinecite{Topping:1927vr}. The 5-point fit covers only the initial uptake to 0.5ML, the 9-point fit extends to 1ML.}
	\label{fig1}
\end{figure}

Using the contact potential difference, the work function of the Kelvin probe, the measured C 1s binding energy and the known C 1s core level binding energy relative to the valence band maximum (283.9 eV), the workfunction, valence band maximum, and conduction band minimum can be determined\cite{Cui:1998br,Maier:2001bl} via the relations:

\begin{align}
\Phi_{D} &= \Phi_{P} + CPD\\
E_{F} - E_{VBM} &= E_{C1s,bulk} - 283.9\mbox{ eV}\\
\chi &= \Phi_{D} + (E_{F} - E_{VBM}) - E_{G}
\end{align}

where $\Phi_{D}$ is the sample workfunction, $\Phi_{P}$ is the probe workfunction and $\chi$ is the electron affinity. Figure \ref{fig1} shows the measured workfunction, electron affinity and $E_{F}-E_{VBM}$ as a function of Mg coverage. The trend is a classic uptake curve\cite{Ibach:2006wc, Kroger:2000tm} reaching a minimum at 0.5 ML and then gradually increasing asymptotically towards the workfunction value expected for bulk Mg ($\approx 3.6$ eV). Changes in the band-bending are minimal, hence the electron affinity essentially tracks the workfunction. The minimum electron affinity is ($-2.01\pm0.05)$ eV at a QCM coverage of 1.5 \AA~(0.5 ML), the workfunction at that coverage being $(2.40\pm0.05)$ eV. A Topping model\cite{Topping:1927vr} fit to the initial uptake region of the workfunction ($0-0.5$ ML) is shown in Figure \ref{fig1}(b) along with a fit to a slightly extended region ($0-1$ ML) to give an estimate of the model parameter sensitivity. The Topping model is given by:

\begin{equation}
\Delta\Phi = -e\frac{p_z}{\varepsilon_0}n_{s}\Theta\left(1 + 8.9\alpha_{e}n_{s}^{3/2}\Theta^{3/2}\right)
\end{equation}

where $\Delta\Phi$ is the change in workfunction, $p_z$ is the dipole moment perpendicular to the surface for an isolated adsorbate complex, $n_s$ is the surface site density, $\alpha_e$ is the polarisability of the dipole complex and $\Theta$ is the adsorbate fractional coverage. As the coverage increases, the dipoles increasingly interact leading to a reduction of the net surface dipole (depolarisation). In the present case, the fits give an initial perpendicular dipole moment of $0.43-0.49$ e\AA~and a surface complex polarizability of $5.0-7.1$ \AA$^{3}$. Assuming a charge separation of approximately 0.75 \AA~normal to the surface,\cite{ODonnell:2015iu} this dipole moment amounts to charge transfer of $0.57-0.65$ electrons per dipole, consistent with the partially ionic bond expected for Mg-O-C complexes on the diamond surface based on earlier theoretical work.

The appearance of a minimum at 0.5ML is consistent with the theoretical model for Mg adsorption on oxygen-terminated diamond\cite{ODonnell:2015iu} and suggests the surface is initially fully oxygen-terminated. The immediate appearance of negative electron affinity with minimal Mg adsorption is, however, quite distinct from the case of Li adsorption where annealing (thermal activation) was required.\cite{ODonnell:2013el}

\begin{figure}[htb]
	\includegraphics[width=0.75\columnwidth]{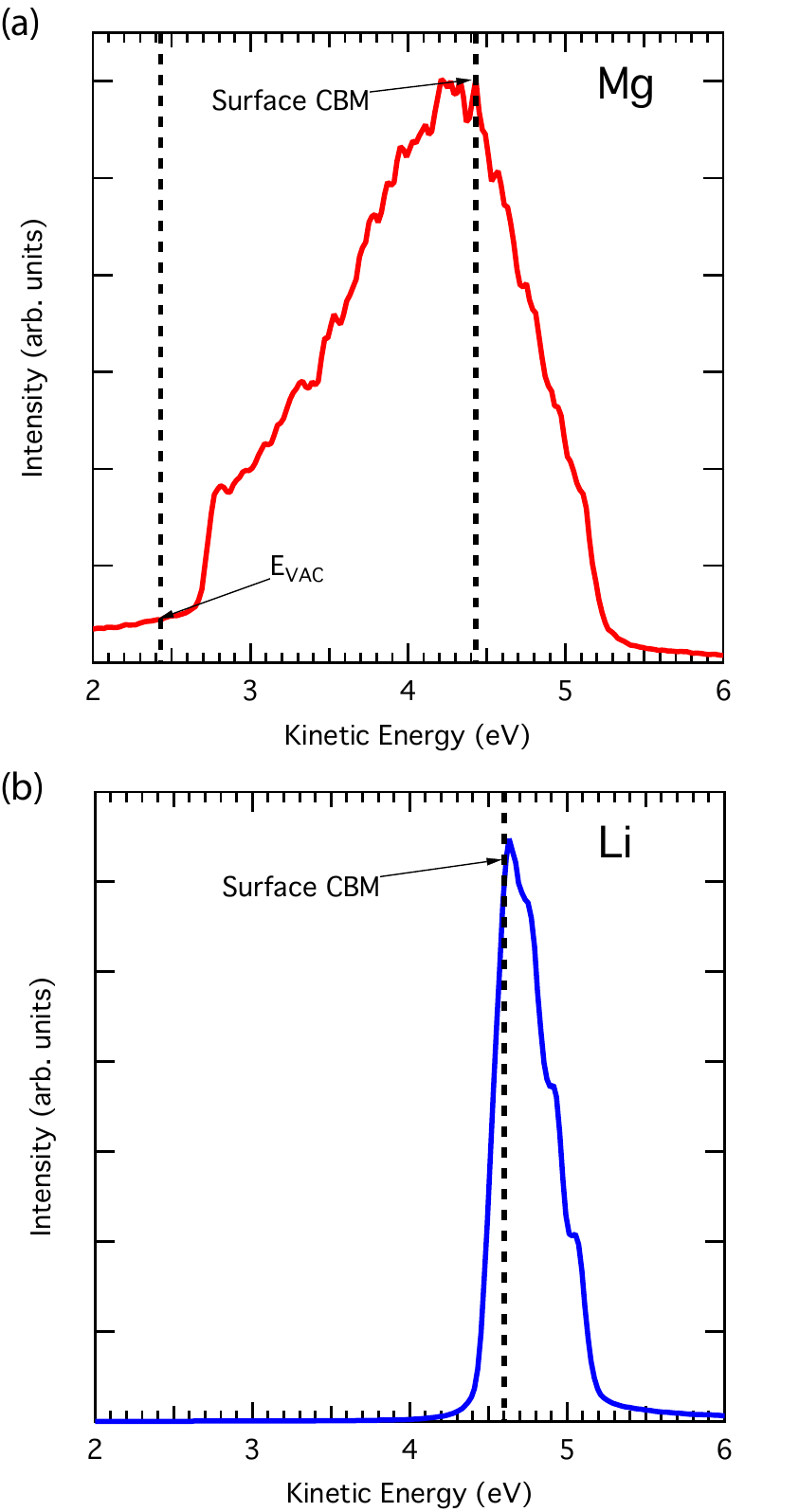}
	\caption{(a) Low kinetic emission spectrum for 0.5ML Mg adsorbed on oxygen-terminated diamond. Vacuum level and surface CBM position, determined by Kelvin probe and XPS C 1s position respectively, are indicated with vertical dashed lines. (b) For comparison, the low kinetic energy emission spectrum for lithium deposited on oxygen-terminated diamond after high temperature activation, from Ref \onlinecite{kanetemp:2014}.}
	\label{fig2}
\end{figure}

Examining the photoemission spectrum of low kinetic energy electrons highlights a second key difference between Li and Mg adsorption. Figure \ref{fig2}(a) shows the low KE spectrum for a 0.5ML Mg deposited surface acquired using a photon energy of 100 eV. A comparison to a thermally activated, lithium-adsorbed surface appears in Figure \ref{fig2}(b) (from Ref \onlinecite{kanetemp:2014}). In Figure \ref{fig2}(a), the energy position of the vacuum level relative to the Fermi level deduced from the Kelvin probe measurements (that is, the workfunction) is shown as a vertical dashed line. The presence of significant emission below the CBM in the case of Mg adsorption is expected based on the very large NEA, allowing emission from above the vacuum level after electrons have thermalized from the CBM into unoccupied surface states lying in the band gap. We have previously shown that this effect on hydrogen-terminated diamond is accompanied by phonon emission during thermalisation and we expect the same to be the case here.\cite{kanetemp:2014} The fact that below CBM emission is apparently absent in the case of lithium-adsorbed diamond is likely due to the NEA for that surface being very small. Next, we draw the readers attention to the emission above the surface CBM. For both surfaces, there is a broad tail of emission extending $0.8-1.0$ eV beyond the surface CBM position. However, for the Mg-absorbed surface, the characteristic oscillations above the CBM seen for Li-adsorbed diamond are missing or at least less distinct. This is a critical observation: we have previously explained these oscillations as emission of electrons that have lost quanta of LO phonon energy as they traverse ballistically from the bulk CBM through the band-bending region towards the surface. Such a model does not depend on the details of the surface chemistry but rather only on two features. First, the band-bending region must be sufficiently narrow. This is clearly the case in both Figures \ref{fig2}(a) and (b) given the broad tail above the CBM. Second, the lateral variation of the electron affinity on the NEA portions of the surface must be sufficiently small on the scale of the LO phonon energy (160 meV) such that the oscillations are not washed out. We propose that such variations explain the lack of oscillations on the Mg-adsorbed surface and how this is related to the absence of thermal activation.

\begin{figure*}[htb]
	\includegraphics[width=0.75\textwidth]{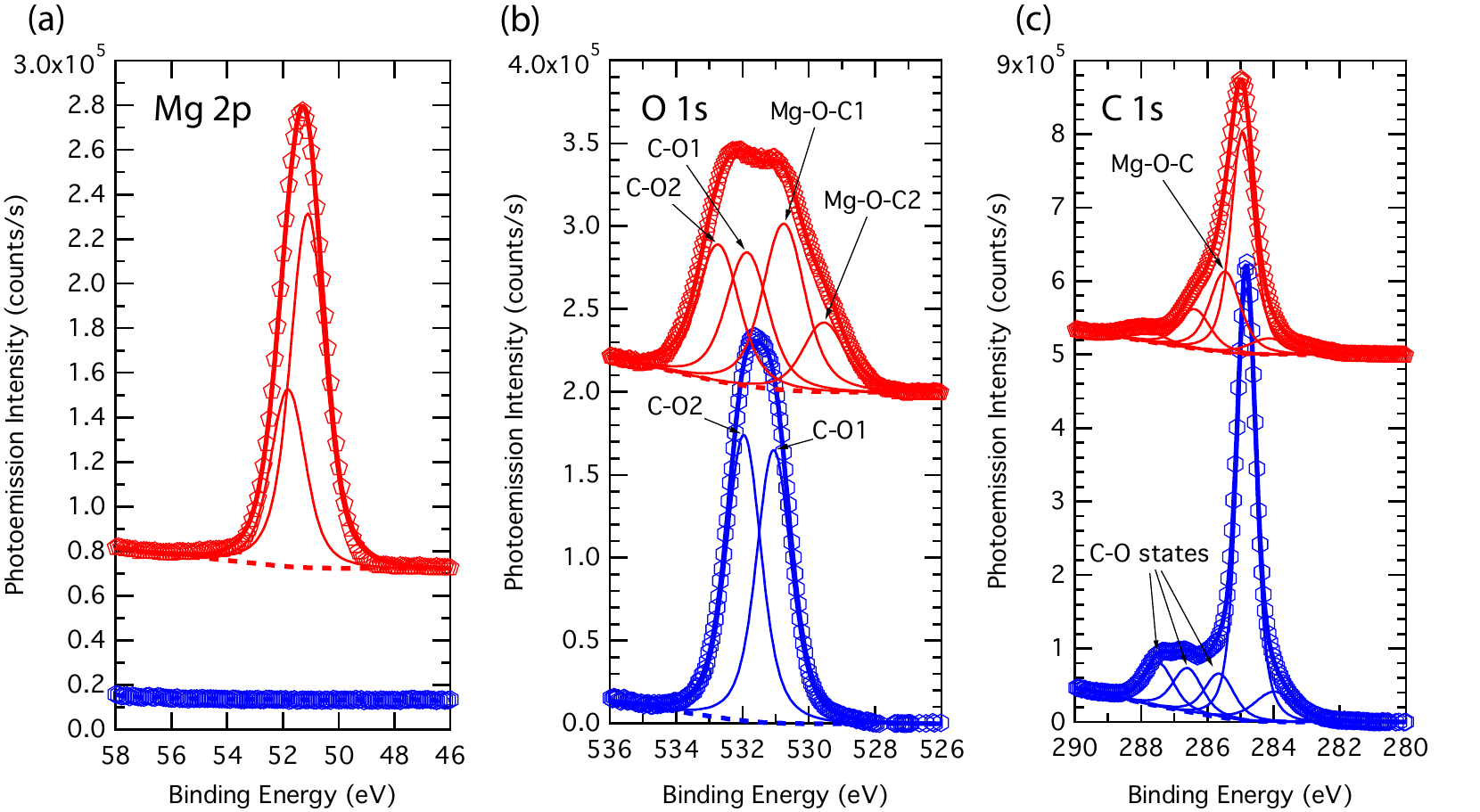}
	\caption{(a) Mg 2p, (b) O 1s and (c) C 1s core level spectra before (blue) and after (red) adsorption of 0.5ML Mg. The surface dipole alters after Mg adsorption, indicated by the shift in C-O components to higher binding energy in the O 1s spectrum simultaneously with a shift in surface components to lower binding energy in the C 1s spectrum. Spectra after absorption have been offset for clarity.}
	\label{fig3}
\end{figure*}

That annealing is not required to induce NEA on Mg-adsorbed C(100):O suggests that Mg-O-C complexes are formed immediately upon Mg adsorption. However, this inevitably means that the random nature of the raindrop deposition model leads to small lateral variations in the electron affinity, washing out the oscillations above the CBM. Variation in electron affinity is not accompanied by a significant variation in $E_{F}-E_{VBM}$ across the surface since this would lead to broadening in the C 1s bulk peak that is not observed here. Consequently, changes in electron affinity must be due to local variations in the surface dipole. Considering the large change in average surface dipole potential even with very low coverages of Mg (Figure \ref{fig1}), it is reasonable to infer that randomly distributed adsorbates give rise to such small variations in the local dipole potential. Such a variation is not observed for the NEA surface induced by Li adsorption and annealing; the hypothesis in that case is that annealing creates small domains of well-ordered Li-O-C complexes. The remainder of the surface in that case is essentially ordinary oxygen-terminated diamond exhibiting PEA and hence not contributing to the low kinetic energy spectrum. 

The observation that the adsorption kinetics for Mg are different from Li is further supported by photoelectron core level spectra. Figure \ref{fig3} shows Mg 2p, O 1s and C 1s core level spectra for the clean and optimally Mg adsorbed surfaces taken using photon energies of 100 eV, 570 eV and 330 eV, respectively, in order to maximize surface sensitivity. After adsorption the Mg 2p spectrum exhibits a single doublet with the 2p 3/2 component at 51.1 eV consistent with charge transfer from Mg to O and giving no evidence of metallic Mg; this is consistent with `raindrop' deposition whereby all Mg atoms interact with the oxygen-terminated diamond substrate and not with each other. In the oxygen core level spectrum, the clean surface initially exhibits two approximately equal peaks at 531.1 and 532.0 eV (features C-O1 and C-O2 in Figure \ref{fig3}). After absorption these components shift to higher binding energies consistent with a change in surface dipole and new peaks at 530.7 and 529.5 eV appear (peaks Mg-O-C1 and Mg-O-C2). The appearance of these new O 1s peaks concurrently with the change in the surface components of the C 1s spectrum after Mg adsorption suggests that the Mg primarily forms Mg-O-C complexes (as expected from theoretical work) rather than a MgO phase independent of the substrate. After absorption, peaks in the O 1s spectrum assigned to Mg-O bonding contribute approximately 50\% of the total spectrum in good agreement with the expected coverage of 0.5ML. We note that these changes in surface chemistry after Mg adsorption are analogous to those observed for Li adsorption after high temperature annealing.\cite{ODonnell:2013el}

\begin{figure}[htb]
	\includegraphics[width=0.75\columnwidth]{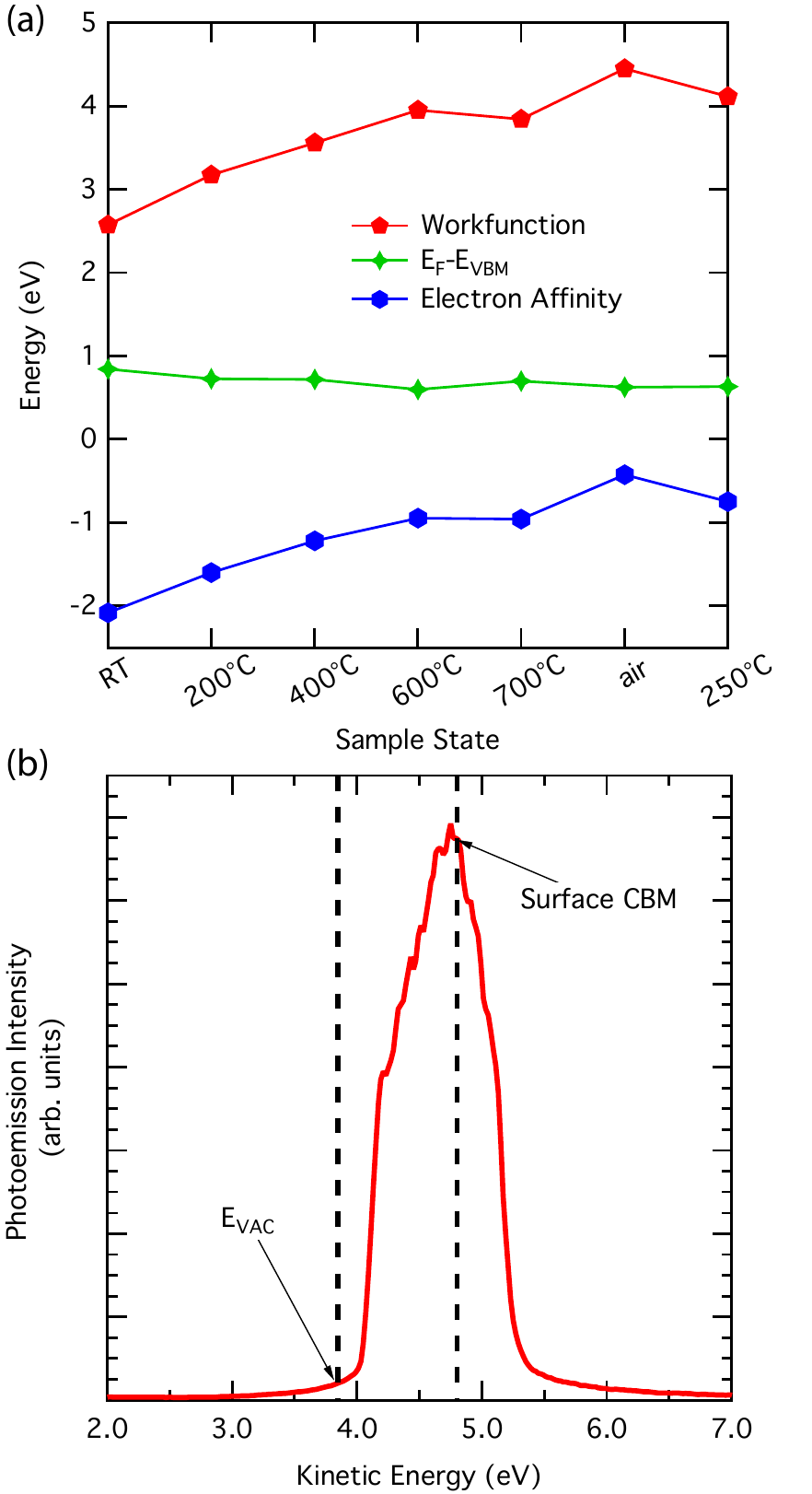}
	\caption{(a): Electron affinity, workfunction and $E_{F}-E_{VBM}$ as a function of annealing temperature, after atmospheric exposure and finally re-annealed in UHV. (b) Low-kinetic energy electron spectrum (photon energy 100 eV) for 0.5 ML Mg adsorption after 700$^{\circ}$C annealing for 15 minutes.}
	\label{fig4}
\end{figure}

Annealing changes the electron affinity and workfunction. Figure \ref{fig4}(a) shows the change in electron affinity, workfunction and $E_{F}-E_{VBM}$ starting with a second diamond sample prepared in an identical manner as the first, with 0.5 ML Mg deposited and subsequently annealing at temperatures of 200, 400, 600 and 700$^{\circ}$C for 15 minutes. The initial electron affinity after deposition is approximately -2.0 eV as with the first sample, demonstrating good reproducibility when starting from a oxygen-plasma termination. The electron affinity increases to approximately -0.9 eV after the final anneal at 700$^{\circ}$C. Annealing is accompanied by a slight loss of both Mg and O consistent with the known tendency of certain oxygen functional groups to leave the diamond surface at relatively low annealing temperatures.\cite{Bobrov:2002tq} We therefore ascribe the change in electron affinity and workfunction with annealing to a reduction in the surface dipole density due to loss of some portion of the surface adsorbates/termination. This behaviour on annealing is distinct from lithium where high temperature annealing activates the surface to a state of negative electron affinity. 

One might hypothesize that annealing the Mg-adsorbed surface might lead to the formation of well-ordered domains similar to that proposed for Li. However, there is no evidence for this in the low kinetic energy electron spectrum [Figure \ref{fig4}(b)]. The spectrum shows a change in vacuum cutoff consistent with the altered workfunction but does not show oscillations above the CBM. This is consistent with a model where a single Mg atom forms two strong bonds with surface oxygen atoms and the resulting complex is fixed and stable with a large local surface dipole. In contrast we can infer that a single Li atom, although forming strong surface bonds (as evidenced by it remaining on the surface despite high temperature annealing) does not form well-incorporated dipole complexes immediately upon adsorption and multiple co-located Li-O-C bonds are instead necessary. The net result is that although theory predicts the ground-state chemistry for Li and Mg adsorption on diamond to be very similar, it is clear experimentally that the kinetics are different. Theoretical work considering isolated Li and Mg atoms on the oxygen-terminated diamond surface is required to elucidate further details. Experimentally, if sufficiently flat oxygen-terminated diamond were available for the task, the Mg and Li systems appear ripe for study with scanning tunneling microscopy where diffusion and domain formation might be observed directly.

Next, we turn to the effect of air exposure and water immersion. Figure \ref{fig4}(a) shows data corresponding to the annealed sample after atmospheric exposure and subsequently re-annealed in UHV at 250$^{\circ}$C to remove atmospheric contaminants (e.g. water). Even after air exposure the sample exhibits a negative electron affinity, with slight reduction likely due to the depolarizing effect of adsorbed water molecules. Re-annealing to remove these absorbed molecules mostly returns the electron affinity to the value prior to atmospheric exposure. Key here is that $E_F - E_{VBM}$ does not substantially change after air exposure: this contrasts with the case of hydrogen-terminated diamond, where a shift of approximately 0.6-0.7 eV in $E_F - E_{VBM}$ is observed following atmospheric exposure.\cite{Edmonds:2011jg} Further, in the case of hydrogen-terminated diamond the sign of $E_F - E_{VBM}$ changes such that the Fermi level lies within the valence band at the surface - such an effect is not observed here, and we conclude that neither upwards band bending nor a surface hole accumulation layer are induced by atmospheric exposure of Mg-adsorbed diamond. The absence of transfer doping despite the considerable NEA suggests there is a sufficiently high occupied surface state density to accommodate holes created by the electrochemical transfer doping reaction. Consequently, few or no mobile holes in the valence band will be generated leading to the absence of surface conductivity. 

\begin{figure}[htb]
	\includegraphics[width=0.75\columnwidth]{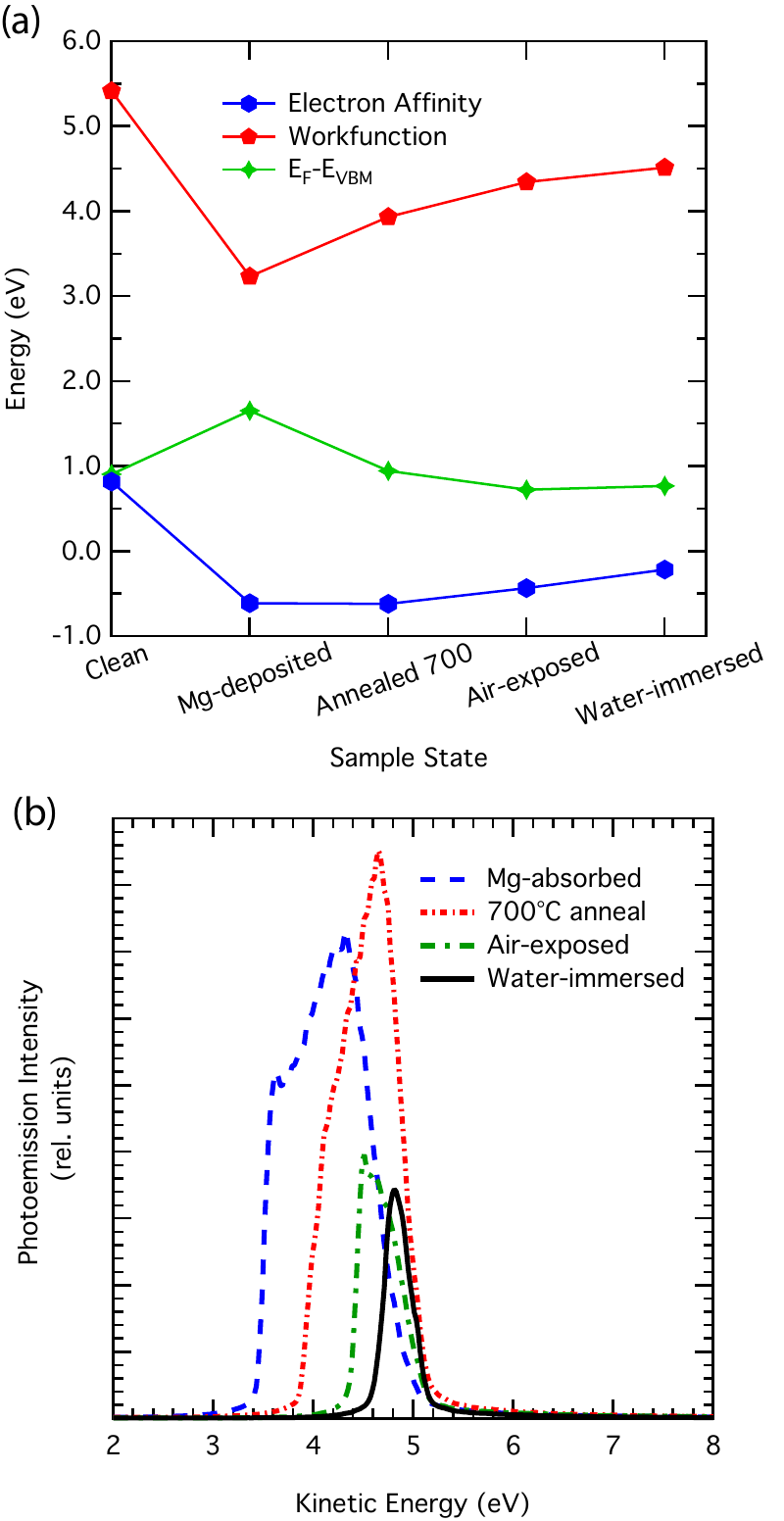}
	\caption{(a) Change in electronic parameters as a function of exposure to air and water for a Mg-absorbed sample. (b) Low kinetic energy electron emission spectra (photon energy 100 eV) for the freshly deposited surface, annealed, air-exposed surface and water immersed surface.}
	\label{fig5}
\end{figure}

Although our largest NEA value was achieved using oxygen-plasma termination, we find it is not necessary to start with this process. Figure \ref{fig5}(a) shows the variation of electron affinity, workfunction and $E_{F}-E_{VBM}$ for a sample initially prepared using a sulphuric/nitric acid step and then deposited with 0.5 ML Mg. The sample was measured before being removed from UHV conditions and left in ambient conditions for 12 hours. After air exposure the sample was re-inserted into UHV and immediately measured. Finally, the sample was removed from UHV, immersed in Milli-Q water for 30 seconds and then returned to UHV and measured. The acid termination method leads to a slightly different behaviour upon Mg absorption: it can be seen from Figure \ref{fig5}(a) that initially the surface band-bending is increased to 1.7 eV. We hypothesize that acid termination leads to surface sites that are not amenable to immediate formation of Mg-O-C complexes upon absorption such as OH groups or even isolated hydrogen-terminated sites. The increase in $E_{F}-E_{VBM}$ is then consistent with gap states induced by metallic species well above the valence band edge. Annealing reverses this effect consistent with the loss of Mg observed in XPS.

The low kinetic energy electron emission spectrum for the air-exposed sample [Figure \ref{fig5}(b)] shows a large emission peak despite the inevitable attenuation due to the presence of adsorbates. This is in contrast to the hydrogen-terminated diamond surface where a reduction in the electron yield by a factor of up to 200 is observed upon air exposure due to the upward band bending resulting from the hole accumulation layer.\cite{Takeuchi:2003jr} Water immersion results in loss of most Mg from the surface ($> 95$\% decrease as determined by the Mg core level) accompanied by an increase in the electron affinity to -0.2 eV. Importantly, the electron yield is only slightly affected, showing the significant influence of even the smallest NEA in removing the emission barrier. The small NEA prevents the substantial emission from below the CBM seen for the fresh Mg-adsorbed surface and for hydrogen-terminated diamond\cite{kanetemp:2014} leading to rather narrow linewidth (0.3 eV FWHM) useful for some electron emission applications. 

For the perfect crystalline Mg-adsorbed O-terminated C(100) surface, one might expect that solvation of surface Mg would be sterically hindered based on the shallow protrusion of Mg sites above the oxygen-terminated surface plane\cite{ODonnell:2015iu} and the reasonably large number of water molecules in the first solvation shell (n=6, Ref \onlinecite{Matwiyoff:1968wm}). However, here we assume again that the oxygen-terminated surfaces in this study are quite rough at the atomic scale, reducing the barrier to solvation. The fact that so little Mg is required to induce at least small NEA makes it more likely that simple in-situ regeneration is possible even in hostile emission environments (e.g. in solution\cite{DiZhu:2013bb}).

\section{Conclusions}

We have demonstrated a negative electron affinity of -2.0 eV on plasma oxidised diamond after Mg absorption. As far as we are aware this is the largest reported NEA on diamond to date. Unlike the lithiated diamond of our previous work, the Mg-absorbed surface does not require a thermal activation process to induce NEA. In view of the expectation of similar ground-state chemistry, we explain the difference between the two systems as arising from differences in adsorption behaviour. That is, magnesium atoms are able to form Mg-O-C complexes with their associated dipole immediately upon adsorption due to the ability to strongly coordinate with two oxygen atoms. In contrast, lithium atoms require thermal energy to distribute on the surface and/or incorporate and form the analogous complexes. Like lithiated diamond however, the Mg-absorbed system is resistant to surface transfer doping and continues to possess a high electron yield even after air exposure. The activation-free process for inducing NEA coupled with the convenience of working with Mg instead of Li positions this surface system as ideal for applications requiring high electron yield outside an ultra-high vacuum environment.

\begin{acknowledgments}

This research was undertaken on the Soft X-ray Spectroscopy beamline at the Australian Synchrotron, Victoria, Australia. M.T.E. is supported by the ARC Laureate Fellowship project (FL120100038) of M.S. Fuhrer. The authors would like to thank Bruce Cowie for assistance at the beamline and Thomas Becker for acquiring atomic force microscopy images of the diamond substrates. This work was performed in part at the Melbourne Centre for Nanofabrication (MCN) in the Victorian Node of the Australian National Fabrication Facility (ANFF).

\end{acknowledgments}


\begin{thebibliography}{27}%
\makeatletter
\providecommand \@ifxundefined [1]{%
 \@ifx{#1\undefined}
}%
\providecommand \@ifnum [1]{%
 \ifnum #1\expandafter \@firstoftwo
 \else \expandafter \@secondoftwo
 \fi
}%
\providecommand \@ifx [1]{%
 \ifx #1\expandafter \@firstoftwo
 \else \expandafter \@secondoftwo
 \fi
}%
\providecommand \natexlab [1]{#1}%
\providecommand \enquote  [1]{``#1''}%
\providecommand \bibnamefont  [1]{#1}%
\providecommand \bibfnamefont [1]{#1}%
\providecommand \citenamefont [1]{#1}%
\providecommand \href@noop [0]{\@secondoftwo}%
\providecommand \href [0]{\begingroup \@sanitize@url \@href}%
\providecommand \@href[1]{\@@startlink{#1}\@@href}%
\providecommand \@@href[1]{\endgroup#1\@@endlink}%
\providecommand \@sanitize@url [0]{\catcode `\\12\catcode `\$12\catcode
  `\&12\catcode `\#12\catcode `\^12\catcode `\_12\catcode `\%12\relax}%
\providecommand \@@startlink[1]{}%
\providecommand \@@endlink[0]{}%
\providecommand \url  [0]{\begingroup\@sanitize@url \@url }%
\providecommand \@url [1]{\endgroup\@href {#1}{\urlprefix }}%
\providecommand \urlprefix  [0]{URL }%
\providecommand \Eprint [0]{\href }%
\providecommand \doibase [0]{http://dx.doi.org/}%
\providecommand \selectlanguage [0]{\@gobble}%
\providecommand \bibinfo  [0]{\@secondoftwo}%
\providecommand \bibfield  [0]{\@secondoftwo}%
\providecommand \translation [1]{[#1]}%
\providecommand \BibitemOpen [0]{}%
\providecommand \bibitemStop [0]{}%
\providecommand \bibitemNoStop [0]{.\EOS\space}%
\providecommand \EOS [0]{\spacefactor3000\relax}%
\providecommand \BibitemShut  [1]{\csname bibitem#1\endcsname}%
\let\auto@bib@innerbib\@empty
\bibitem [{\citenamefont {Himpsel}\ \emph {et~al.}(1979)\citenamefont
  {Himpsel}, \citenamefont {Knapp}, \citenamefont {VanVechten},\ and\
  \citenamefont {Eastman}}]{Himpsel:1979vd}%
  \BibitemOpen
  \bibfield  {author} {\bibinfo {author} {\bibfnamefont {F.}~\bibnamefont
  {Himpsel}}, \bibinfo {author} {\bibfnamefont {J.}~\bibnamefont {Knapp}},
  \bibinfo {author} {\bibfnamefont {J.}~\bibnamefont {VanVechten}}, \ and\
  \bibinfo {author} {\bibfnamefont {D.}~\bibnamefont {Eastman}},\ }\href@noop
  {} {\bibfield  {journal} {\bibinfo  {journal} {Physical Review B: Condensed
  Matter}\ }\textbf {\bibinfo {volume} {20}},\ \bibinfo {pages} {624} (\bibinfo
  {year} {1979})}\BibitemShut {NoStop}%
\bibitem [{\citenamefont {O'Donnell}\ \emph {et~al.}(2013)\citenamefont
  {O'Donnell}, \citenamefont {Edmonds}, \citenamefont {Ristein}, \citenamefont
  {Tadich}, \citenamefont {Thomsen}, \citenamefont {Wu}, \citenamefont
  {Pakes},\ and\ \citenamefont {Ley}}]{ODonnell:2013el}%
  \BibitemOpen
  \bibfield  {author} {\bibinfo {author} {\bibfnamefont {K.~M.}\ \bibnamefont
  {O'Donnell}}, \bibinfo {author} {\bibfnamefont {M.~T.}\ \bibnamefont
  {Edmonds}}, \bibinfo {author} {\bibfnamefont {J.}~\bibnamefont {Ristein}},
  \bibinfo {author} {\bibfnamefont {A.}~\bibnamefont {Tadich}}, \bibinfo
  {author} {\bibfnamefont {L.}~\bibnamefont {Thomsen}}, \bibinfo {author}
  {\bibfnamefont {Q.-H.}\ \bibnamefont {Wu}}, \bibinfo {author} {\bibfnamefont
  {C.~I.}\ \bibnamefont {Pakes}}, \ and\ \bibinfo {author} {\bibfnamefont
  {L.}~\bibnamefont {Ley}},\ }\href@noop {} {\bibfield  {journal} {\bibinfo
  {journal} {Advanced Functional Materials}\ }\textbf {\bibinfo {volume}
  {23}},\ \bibinfo {pages} {5608} (\bibinfo {year} {2013})}\BibitemShut
  {NoStop}%
\bibitem [{\citenamefont {Shih}\ \emph {et~al.}(1997)\citenamefont {Shih},
  \citenamefont {Yater}, \citenamefont {Pehrsson}, \citenamefont {Butler},
  \citenamefont {Hor},\ and\ \citenamefont {Abrams}}]{Shih:1997bt}%
  \BibitemOpen
  \bibfield  {author} {\bibinfo {author} {\bibfnamefont {A.}~\bibnamefont
  {Shih}}, \bibinfo {author} {\bibfnamefont {J.}~\bibnamefont {Yater}},
  \bibinfo {author} {\bibfnamefont {P.}~\bibnamefont {Pehrsson}}, \bibinfo
  {author} {\bibfnamefont {J.}~\bibnamefont {Butler}}, \bibinfo {author}
  {\bibfnamefont {C.}~\bibnamefont {Hor}}, \ and\ \bibinfo {author}
  {\bibfnamefont {R.}~\bibnamefont {Abrams}},\ }\href@noop {} {\bibfield
  {journal} {\bibinfo  {journal} {Journal of Applied Physics}\ }\textbf
  {\bibinfo {volume} {82}},\ \bibinfo {pages} {1860} (\bibinfo {year}
  {1997})}\BibitemShut {NoStop}%
\bibitem [{\citenamefont {Sun}\ \emph {et~al.}(2014)\citenamefont {Sun},
  \citenamefont {Koeck}, \citenamefont {Rezikyan}, \citenamefont {Treacy},\
  and\ \citenamefont {Nemanich}}]{Sun:2014ih}%
  \BibitemOpen
  \bibfield  {author} {\bibinfo {author} {\bibfnamefont {T.}~\bibnamefont
  {Sun}}, \bibinfo {author} {\bibfnamefont {F.~A.~M.}\ \bibnamefont {Koeck}},
  \bibinfo {author} {\bibfnamefont {A.}~\bibnamefont {Rezikyan}}, \bibinfo
  {author} {\bibfnamefont {M.~M.~J.}\ \bibnamefont {Treacy}}, \ and\ \bibinfo
  {author} {\bibfnamefont {R.~J.}\ \bibnamefont {Nemanich}},\ }\href@noop {}
  {\bibfield  {journal} {\bibinfo  {journal} {Physical Review B: Condensed
  Matter}\ }\textbf {\bibinfo {volume} {90}},\ \bibinfo {pages} {121302}
  (\bibinfo {year} {2014})}\BibitemShut {NoStop}%
\bibitem [{\citenamefont {Chang}\ \emph {et~al.}(2010)\citenamefont {Chang},
  \citenamefont {Wu}, \citenamefont {Ben-Zvi}, \citenamefont {Burrill},
  \citenamefont {Kewisch}, \citenamefont {Rao}, \citenamefont {Smedley},
  \citenamefont {Wang}, \citenamefont {Muller}, \citenamefont {Busby},\ and\
  \citenamefont {Dimitrov}}]{Chang:2010fa}%
  \BibitemOpen
  \bibfield  {author} {\bibinfo {author} {\bibfnamefont {X.}~\bibnamefont
  {Chang}}, \bibinfo {author} {\bibfnamefont {Q.}~\bibnamefont {Wu}}, \bibinfo
  {author} {\bibfnamefont {I.}~\bibnamefont {Ben-Zvi}}, \bibinfo {author}
  {\bibfnamefont {A.}~\bibnamefont {Burrill}}, \bibinfo {author} {\bibfnamefont
  {J.}~\bibnamefont {Kewisch}}, \bibinfo {author} {\bibfnamefont
  {T.}~\bibnamefont {Rao}}, \bibinfo {author} {\bibfnamefont {J.}~\bibnamefont
  {Smedley}}, \bibinfo {author} {\bibfnamefont {E.}~\bibnamefont {Wang}},
  \bibinfo {author} {\bibfnamefont {E.}~\bibnamefont {Muller}}, \bibinfo
  {author} {\bibfnamefont {R.}~\bibnamefont {Busby}}, \ and\ \bibinfo {author}
  {\bibfnamefont {D.}~\bibnamefont {Dimitrov}},\ }\href@noop {} {\bibfield
  {journal} {\bibinfo  {journal} {Physical Review Letters}\ }\textbf {\bibinfo
  {volume} {105}},\ \bibinfo {pages} {164801} (\bibinfo {year}
  {2010})}\BibitemShut {NoStop}%
\bibitem [{\citenamefont {Zhu}\ \emph {et~al.}(2013)\citenamefont {Zhu},
  \citenamefont {Zhang}, \citenamefont {Ruther},\ and\ \citenamefont
  {Hamers}}]{DiZhu:2013bb}%
  \BibitemOpen
  \bibfield  {author} {\bibinfo {author} {\bibfnamefont {D.}~\bibnamefont
  {Zhu}}, \bibinfo {author} {\bibfnamefont {L.}~\bibnamefont {Zhang}}, \bibinfo
  {author} {\bibfnamefont {R.~E.}\ \bibnamefont {Ruther}}, \ and\ \bibinfo
  {author} {\bibfnamefont {R.~J.}\ \bibnamefont {Hamers}},\ }\href@noop {}
  {\bibfield  {journal} {\bibinfo  {journal} {Nature Materials}\ }\textbf
  {\bibinfo {volume} {12}},\ \bibinfo {pages} {836} (\bibinfo {year}
  {2013})}\BibitemShut {NoStop}%
\bibitem [{\citenamefont {Cui}\ \emph {et~al.}(1998)\citenamefont {Cui},
  \citenamefont {Ristein},\ and\ \citenamefont {Ley}}]{Cui:1998br}%
  \BibitemOpen
  \bibfield  {author} {\bibinfo {author} {\bibfnamefont {J.~B.}\ \bibnamefont
  {Cui}}, \bibinfo {author} {\bibfnamefont {J.}~\bibnamefont {Ristein}}, \ and\
  \bibinfo {author} {\bibfnamefont {L.}~\bibnamefont {Ley}},\ }\href@noop {}
  {\bibfield  {journal} {\bibinfo  {journal} {Physical Review Letters}\
  }\textbf {\bibinfo {volume} {81}},\ \bibinfo {pages} {429} (\bibinfo {year}
  {1998})}\BibitemShut {NoStop}%
\bibitem [{\citenamefont {Graupner}\ \emph {et~al.}(1998)\citenamefont
  {Graupner}, \citenamefont {Maier}, \citenamefont {Ristein}, \citenamefont
  {Ley},\ and\ \citenamefont {Jung}}]{Graupner:1998ub}%
  \BibitemOpen
  \bibfield  {author} {\bibinfo {author} {\bibfnamefont {R.}~\bibnamefont
  {Graupner}}, \bibinfo {author} {\bibfnamefont {F.}~\bibnamefont {Maier}},
  \bibinfo {author} {\bibfnamefont {J.}~\bibnamefont {Ristein}}, \bibinfo
  {author} {\bibfnamefont {L.}~\bibnamefont {Ley}}, \ and\ \bibinfo {author}
  {\bibfnamefont {C.}~\bibnamefont {Jung}},\ }\href@noop {} {\bibfield
  {journal} {\bibinfo  {journal} {Physical Review B: Condensed Matter}\
  }\textbf {\bibinfo {volume} {57}},\ \bibinfo {pages} {12397} (\bibinfo {year}
  {1998})}\BibitemShut {NoStop}%
\bibitem [{\citenamefont {Graupner}\ \emph {et~al.}(1999)\citenamefont
  {Graupner}, \citenamefont {Ristein}, \citenamefont {Ley},\ and\ \citenamefont
  {Jung}}]{Graupner:1999up}%
  \BibitemOpen
  \bibfield  {author} {\bibinfo {author} {\bibfnamefont {R.}~\bibnamefont
  {Graupner}}, \bibinfo {author} {\bibfnamefont {J.}~\bibnamefont {Ristein}},
  \bibinfo {author} {\bibfnamefont {L.}~\bibnamefont {Ley}}, \ and\ \bibinfo
  {author} {\bibfnamefont {C.}~\bibnamefont {Jung}},\ }\href@noop {} {\bibfield
   {journal} {\bibinfo  {journal} {Physical Review B: Condensed Matter}\
  }\textbf {\bibinfo {volume} {60}},\ \bibinfo {pages} {17023} (\bibinfo {year}
  {1999})}\BibitemShut {NoStop}%
\bibitem [{\citenamefont {Van~der Weide}\ \emph {et~al.}(1994)\citenamefont
  {Van~der Weide}, \citenamefont {Zhang}, \citenamefont {Baumann},
  \citenamefont {Wensell}, \citenamefont {Bernholc},\ and\ \citenamefont
  {Nemanich}}]{VanderWeide:1994wz}%
  \BibitemOpen
  \bibfield  {author} {\bibinfo {author} {\bibfnamefont {J.}~\bibnamefont
  {Van~der Weide}}, \bibinfo {author} {\bibfnamefont {Z.}~\bibnamefont
  {Zhang}}, \bibinfo {author} {\bibfnamefont {P.}~\bibnamefont {Baumann}},
  \bibinfo {author} {\bibfnamefont {M.}~\bibnamefont {Wensell}}, \bibinfo
  {author} {\bibfnamefont {J.}~\bibnamefont {Bernholc}}, \ and\ \bibinfo
  {author} {\bibfnamefont {R.~J.}\ \bibnamefont {Nemanich}},\ }\href@noop {}
  {\bibfield  {journal} {\bibinfo  {journal} {Physical Review B: Condensed
  Matter}\ }\textbf {\bibinfo {volume} {50}},\ \bibinfo {pages} {5803}
  (\bibinfo {year} {1994})}\BibitemShut {NoStop}%
\bibitem [{\citenamefont {O'Donnell}\ \emph
  {et~al.}(2014{\natexlab{a}})\citenamefont {O'Donnell}, \citenamefont
  {Edmonds}, \citenamefont {Ristein}, \citenamefont {Rietwyk}, \citenamefont
  {Tadich}, \citenamefont {Thomsen}, \citenamefont {Pakes},\ and\ \citenamefont
  {Ley}}]{kanetemp:2014}%
  \BibitemOpen
  \bibfield  {author} {\bibinfo {author} {\bibfnamefont {K.~M.}\ \bibnamefont
  {O'Donnell}}, \bibinfo {author} {\bibfnamefont {M.~T.}\ \bibnamefont
  {Edmonds}}, \bibinfo {author} {\bibfnamefont {J.}~\bibnamefont {Ristein}},
  \bibinfo {author} {\bibfnamefont {K.~J.}\ \bibnamefont {Rietwyk}}, \bibinfo
  {author} {\bibfnamefont {A.}~\bibnamefont {Tadich}}, \bibinfo {author}
  {\bibfnamefont {L.}~\bibnamefont {Thomsen}}, \bibinfo {author} {\bibfnamefont
  {C.~I.}\ \bibnamefont {Pakes}}, \ and\ \bibinfo {author} {\bibfnamefont
  {L.}~\bibnamefont {Ley}},\ }\href@noop {} {\bibfield  {journal} {\bibinfo
  {journal} {Journal of Physics: Condensed Matter}\ }\textbf {\bibinfo {volume}
  {26}},\ \bibinfo {pages} {395008} (\bibinfo {year}
  {2014}{\natexlab{a}})}\BibitemShut {NoStop}%
\bibitem [{\citenamefont {Maier}\ \emph {et~al.}(2000)\citenamefont {Maier},
  \citenamefont {Riedel}, \citenamefont {Mantel}, \citenamefont {Ristein},\
  and\ \citenamefont {Ley}}]{Maier:2000uq}%
  \BibitemOpen
  \bibfield  {author} {\bibinfo {author} {\bibfnamefont {F.}~\bibnamefont
  {Maier}}, \bibinfo {author} {\bibfnamefont {M.}~\bibnamefont {Riedel}},
  \bibinfo {author} {\bibfnamefont {B.}~\bibnamefont {Mantel}}, \bibinfo
  {author} {\bibfnamefont {J.}~\bibnamefont {Ristein}}, \ and\ \bibinfo
  {author} {\bibfnamefont {L.}~\bibnamefont {Ley}},\ }\href@noop {} {\bibfield
  {journal} {\bibinfo  {journal} {Physical Review Letters}\ }\textbf {\bibinfo
  {volume} {85}},\ \bibinfo {pages} {3472} (\bibinfo {year}
  {2000})}\BibitemShut {NoStop}%
\bibitem [{\citenamefont {Riedel}\ \emph {et~al.}(2004)\citenamefont {Riedel},
  \citenamefont {Ristein},\ and\ \citenamefont {Ley}}]{Riedel:2004gl}%
  \BibitemOpen
  \bibfield  {author} {\bibinfo {author} {\bibfnamefont {M.}~\bibnamefont
  {Riedel}}, \bibinfo {author} {\bibfnamefont {J.}~\bibnamefont {Ristein}}, \
  and\ \bibinfo {author} {\bibfnamefont {L.}~\bibnamefont {Ley}},\ }\href@noop
  {} {\bibfield  {journal} {\bibinfo  {journal} {Physical Review B: Condensed
  Matter}\ }\textbf {\bibinfo {volume} {69}},\ \bibinfo {pages} {125338}
  (\bibinfo {year} {2004})}\BibitemShut {NoStop}%
\bibitem [{\citenamefont {Takeuchi}\ \emph {et~al.}(2003)\citenamefont
  {Takeuchi}, \citenamefont {Riedel}, \citenamefont {Ristein},\ and\
  \citenamefont {Ley}}]{Takeuchi:2003jr}%
  \BibitemOpen
  \bibfield  {author} {\bibinfo {author} {\bibfnamefont {D.}~\bibnamefont
  {Takeuchi}}, \bibinfo {author} {\bibfnamefont {M.}~\bibnamefont {Riedel}},
  \bibinfo {author} {\bibfnamefont {J.}~\bibnamefont {Ristein}}, \ and\
  \bibinfo {author} {\bibfnamefont {L.}~\bibnamefont {Ley}},\ }\href@noop {}
  {\bibfield  {journal} {\bibinfo  {journal} {Physical Review B: Condensed
  Matter}\ }\textbf {\bibinfo {volume} {68}},\ \bibinfo {pages} {041304}
  (\bibinfo {year} {2003})}\BibitemShut {NoStop}%
\bibitem [{\citenamefont {Foord}\ \emph
  {et~al.}(2001{\natexlab{a}})\citenamefont {Foord}, \citenamefont {Wang},
  \citenamefont {Hian~Lau}, \citenamefont {Hiramatsu}, \citenamefont
  {Vickers},\ and\ \citenamefont {Jackman}}]{Foord:2001ux}%
  \BibitemOpen
  \bibfield  {author} {\bibinfo {author} {\bibfnamefont {J.~S.}\ \bibnamefont
  {Foord}}, \bibinfo {author} {\bibfnamefont {J.}~\bibnamefont {Wang}},
  \bibinfo {author} {\bibfnamefont {C.}~\bibnamefont {Hian~Lau}}, \bibinfo
  {author} {\bibfnamefont {M.}~\bibnamefont {Hiramatsu}}, \bibinfo {author}
  {\bibfnamefont {J.}~\bibnamefont {Vickers}}, \ and\ \bibinfo {author}
  {\bibfnamefont {R.~B.}\ \bibnamefont {Jackman}},\ }\href@noop {} {\bibfield
  {journal} {\bibinfo  {journal} {Physica Status Solidi A}\ }\textbf {\bibinfo
  {volume} {186}},\ \bibinfo {pages} {227} (\bibinfo {year}
  {2001}{\natexlab{a}})}\BibitemShut {NoStop}%
\bibitem [{\citenamefont {Foord}\ \emph
  {et~al.}(2001{\natexlab{b}})\citenamefont {Foord}, \citenamefont {Hian},\
  and\ \citenamefont {Jackman}}]{Anonymous:A0ci7tur}%
  \BibitemOpen
  \bibfield  {author} {\bibinfo {author} {\bibfnamefont {J.~S.}\ \bibnamefont
  {Foord}}, \bibinfo {author} {\bibfnamefont {L.~C.}\ \bibnamefont {Hian}}, \
  and\ \bibinfo {author} {\bibfnamefont {R.~B.}\ \bibnamefont {Jackman}},\
  }\href@noop {} {\bibfield  {journal} {\bibinfo  {journal} {Diamond and
  Related Materials}\ }\textbf {\bibinfo {volume} {10}},\ \bibinfo {pages}
  {710} (\bibinfo {year} {2001}{\natexlab{b}})}\BibitemShut {NoStop}%
\bibitem [{\citenamefont {O'Donnell}\ \emph {et~al.}(2010)\citenamefont
  {O'Donnell}, \citenamefont {Martin}, \citenamefont {Fox},\ and\ \citenamefont
  {Cherns}}]{ODonnell:2010fw}%
  \BibitemOpen
  \bibfield  {author} {\bibinfo {author} {\bibfnamefont {K.~M.}\ \bibnamefont
  {O'Donnell}}, \bibinfo {author} {\bibfnamefont {T.}~\bibnamefont {Martin}},
  \bibinfo {author} {\bibfnamefont {N.~A.}\ \bibnamefont {Fox}}, \ and\
  \bibinfo {author} {\bibfnamefont {D.}~\bibnamefont {Cherns}},\ }\href@noop {}
  {\bibfield  {journal} {\bibinfo  {journal} {Physical Review B: Condensed
  Matter}\ }\textbf {\bibinfo {volume} {82}},\ \bibinfo {pages} {115303}
  (\bibinfo {year} {2010})}\BibitemShut {NoStop}%
\bibitem [{\citenamefont {O'Donnell}\ \emph
  {et~al.}(2014{\natexlab{b}})\citenamefont {O'Donnell}, \citenamefont
  {Martin}, \citenamefont {Edmonds}, \citenamefont {Tadich}, \citenamefont
  {Thomsen}, \citenamefont {Ristein}, \citenamefont {Pakes}, \citenamefont
  {Fox},\ and\ \citenamefont {Ley}}]{ODonnell:2014cp}%
  \BibitemOpen
  \bibfield  {author} {\bibinfo {author} {\bibfnamefont {K.~M.}\ \bibnamefont
  {O'Donnell}}, \bibinfo {author} {\bibfnamefont {T.~L.}\ \bibnamefont
  {Martin}}, \bibinfo {author} {\bibfnamefont {M.~T.}\ \bibnamefont {Edmonds}},
  \bibinfo {author} {\bibfnamefont {A.}~\bibnamefont {Tadich}}, \bibinfo
  {author} {\bibfnamefont {L.}~\bibnamefont {Thomsen}}, \bibinfo {author}
  {\bibfnamefont {J.}~\bibnamefont {Ristein}}, \bibinfo {author} {\bibfnamefont
  {C.~I.}\ \bibnamefont {Pakes}}, \bibinfo {author} {\bibfnamefont {N.~A.}\
  \bibnamefont {Fox}}, \ and\ \bibinfo {author} {\bibfnamefont
  {L.}~\bibnamefont {Ley}},\ }\href@noop {} {\bibfield  {journal} {\bibinfo
  {journal} {Physica Status Solidi A}\ }\textbf {\bibinfo {volume} {211}},\
  \bibinfo {pages} {2209} (\bibinfo {year} {2014}{\natexlab{b}})}\BibitemShut
  {NoStop}%
\bibitem [{\citenamefont {O'Donnell}\ \emph {et~al.}(2015)\citenamefont
  {O'Donnell}, \citenamefont {Martin},\ and\ \citenamefont
  {Allan}}]{ODonnell:2015iu}%
  \BibitemOpen
  \bibfield  {author} {\bibinfo {author} {\bibfnamefont {K.~M.}\ \bibnamefont
  {O'Donnell}}, \bibinfo {author} {\bibfnamefont {T.~L.}\ \bibnamefont
  {Martin}}, \ and\ \bibinfo {author} {\bibfnamefont {N.~L.}\ \bibnamefont
  {Allan}},\ }\href@noop {} {\bibfield  {journal} {\bibinfo  {journal} {Chem.
  Mater}\ }\textbf {\bibinfo {volume} {27}},\ \bibinfo {pages} {1306} (\bibinfo
  {year} {2015})}\BibitemShut {NoStop}%
\bibitem [{\citenamefont {Cowie}\ \emph {et~al.}(2010)\citenamefont {Cowie},
  \citenamefont {Tadich},\ and\ \citenamefont {Thomsen}}]{Cowie:2010cm}%
  \BibitemOpen
  \bibfield  {author} {\bibinfo {author} {\bibfnamefont {B.~C.~C.}\
  \bibnamefont {Cowie}}, \bibinfo {author} {\bibfnamefont {A.}~\bibnamefont
  {Tadich}}, \ and\ \bibinfo {author} {\bibfnamefont {L.}~\bibnamefont
  {Thomsen}},\ }\href@noop {} {\bibfield  {journal} {\bibinfo  {journal} {AIP
  Conference Proceedings}\ }\textbf {\bibinfo {volume} {1234}},\ \bibinfo
  {pages} {307} (\bibinfo {year} {2010})}\BibitemShut {NoStop}%
\bibitem [{\citenamefont {Maier}\ \emph {et~al.}(2001)\citenamefont {Maier},
  \citenamefont {Ristein},\ and\ \citenamefont {Ley}}]{Maier:2001bl}%
  \BibitemOpen
  \bibfield  {author} {\bibinfo {author} {\bibfnamefont {F.}~\bibnamefont
  {Maier}}, \bibinfo {author} {\bibfnamefont {J.}~\bibnamefont {Ristein}}, \
  and\ \bibinfo {author} {\bibfnamefont {L.}~\bibnamefont {Ley}},\ }\href@noop
  {} {\bibfield  {journal} {\bibinfo  {journal} {Physical Review B: Condensed
  Matter}\ }\textbf {\bibinfo {volume} {64}},\ \bibinfo {pages} {165411}
  (\bibinfo {year} {2001})}\BibitemShut {NoStop}%
\bibitem [{\citenamefont {Ibach}(2006)}]{Ibach:2006wc}%
  \BibitemOpen
  \bibfield  {author} {\bibinfo {author} {\bibfnamefont {H.}~\bibnamefont
  {Ibach}},\ }\href@noop {} {\emph {\bibinfo {title} {{Physics of Surfaces and
  Interfaces}}}}\ (\bibinfo  {publisher} {Springer},\ \bibinfo {address}
  {Berlin},\ \bibinfo {year} {2006})\BibitemShut {NoStop}%
\bibitem [{\citenamefont {Kroger}\ \emph {et~al.}(2000)\citenamefont {Kroger},
  \citenamefont {Bruchmann}, \citenamefont {Lehwald},\ and\ \citenamefont
  {Ibach}}]{Kroger:2000tm}%
  \BibitemOpen
  \bibfield  {author} {\bibinfo {author} {\bibfnamefont {J.}~\bibnamefont
  {Kroger}}, \bibinfo {author} {\bibfnamefont {D.}~\bibnamefont {Bruchmann}},
  \bibinfo {author} {\bibfnamefont {S.}~\bibnamefont {Lehwald}}, \ and\
  \bibinfo {author} {\bibfnamefont {H.}~\bibnamefont {Ibach}},\ }\href@noop {}
  {\bibfield  {journal} {\bibinfo  {journal} {Surface Science}\ }\textbf
  {\bibinfo {volume} {449}},\ \bibinfo {pages} {227} (\bibinfo {year}
  {2000})}\BibitemShut {NoStop}%
\bibitem [{\citenamefont {Topping}(1927)}]{Topping:1927vr}%
  \BibitemOpen
  \bibfield  {author} {\bibinfo {author} {\bibfnamefont {J.}~\bibnamefont
  {Topping}},\ }\href@noop {} {\bibfield  {journal} {\bibinfo  {journal}
  {Proceedings of the Royal Society of London. Series A}\ }\textbf {\bibinfo
  {volume} {114}},\ \bibinfo {pages} {67} (\bibinfo {year} {1927})}\BibitemShut
  {NoStop}%
\bibitem [{\citenamefont {Bobrov}\ \emph {et~al.}(2002)\citenamefont {Bobrov},
  \citenamefont {Shechter}, \citenamefont {Hoffman},\ and\ \citenamefont
  {Folman}}]{Bobrov:2002tq}%
  \BibitemOpen
  \bibfield  {author} {\bibinfo {author} {\bibfnamefont {K.}~\bibnamefont
  {Bobrov}}, \bibinfo {author} {\bibfnamefont {H.}~\bibnamefont {Shechter}},
  \bibinfo {author} {\bibfnamefont {A.}~\bibnamefont {Hoffman}}, \ and\
  \bibinfo {author} {\bibfnamefont {M.}~\bibnamefont {Folman}},\ }\href@noop {}
  {\bibfield  {journal} {\bibinfo  {journal} {Applications of Surface Science}\
  }\textbf {\bibinfo {volume} {196}},\ \bibinfo {pages} {173} (\bibinfo {year}
  {2002})}\BibitemShut {NoStop}%
\bibitem [{\citenamefont {Edmonds}\ \emph {et~al.}(2011)\citenamefont
  {Edmonds}, \citenamefont {Pakes}, \citenamefont {Mammadov}, \citenamefont
  {Zhang}, \citenamefont {Tadich}, \citenamefont {Ristein},\ and\ \citenamefont
  {Ley}}]{Edmonds:2011jg}%
  \BibitemOpen
  \bibfield  {author} {\bibinfo {author} {\bibfnamefont {M.~T.}\ \bibnamefont
  {Edmonds}}, \bibinfo {author} {\bibfnamefont {C.~I.}\ \bibnamefont {Pakes}},
  \bibinfo {author} {\bibfnamefont {S.}~\bibnamefont {Mammadov}}, \bibinfo
  {author} {\bibfnamefont {W.}~\bibnamefont {Zhang}}, \bibinfo {author}
  {\bibfnamefont {A.}~\bibnamefont {Tadich}}, \bibinfo {author} {\bibfnamefont
  {J.}~\bibnamefont {Ristein}}, \ and\ \bibinfo {author} {\bibfnamefont
  {L.}~\bibnamefont {Ley}},\ }\href@noop {} {\bibfield  {journal} {\bibinfo
  {journal} {Physica Status Solidi A}\ }\textbf {\bibinfo {volume} {208}},\
  \bibinfo {pages} {2062} (\bibinfo {year} {2011})}\BibitemShut {NoStop}%
\bibitem [{\citenamefont {Matwiyoff}\ and\ \citenamefont
  {Taube}(1968)}]{Matwiyoff:1968wm}%
  \BibitemOpen
  \bibfield  {author} {\bibinfo {author} {\bibfnamefont {N.~A.}\ \bibnamefont
  {Matwiyoff}}\ and\ \bibinfo {author} {\bibfnamefont {H.}~\bibnamefont
  {Taube}},\ }\href@noop {} {\bibfield  {journal} {\bibinfo  {journal} {Journal
  of the American Chemical Society}\ }\textbf {\bibinfo {volume} {90}},\
  \bibinfo {pages} {2796} (\bibinfo {year} {1968})}\BibitemShut {NoStop}%
\end{thebibliography}
\end{document}